\documentclass[epj]{svjour}
\usepackage{graphicx}
\newcommand{\eqn}[1]{eqn.~\ref{#1}}
\newcommand{\tsnk}{t_{\rm snk}}
\begin{document}
\title{Nucleon Structure from Lattice QCD
\author{David Richards}
}                     
%
%
\institute{Jefferson Laboratory, 12000 Jefferson Avenue, Newport News, VA 23606, USA}
%
\date{November 12, 2007}
%
\abstract{
Recent advances in lattice field theory, in computer technology and in
chiral perturbation theory have enabled lattice QCD to emerge as a
powerful quantitative tool in understanding hadron structure.  I
describe recent progress in the computation of the nucleon form
factors and moments of parton distribution functions, before
proceeding to describe lattice studies of the Generalized Parton
Distributions (GPDs).  In particular, I show how lattice studies of
GPDs contribute to building a three-dimensional picture of the proton.
I conclude by describing the prospects
for studying the structure of resonances from lattice QCD.
} 
\maketitle
\section{Introduction}
\label{intro}
The ability for lattice QCD to make increasingly precise, \textit{ab
  initio} calculations of hadron structure has been driven by advances
  in three areas: algorithmic advances, enabling both chiral symmetry
  to be more faithfully represented on the lattice and gauge
  configurations to be generated more efficiently, computational
  advances represented by the availability of special-purpose and
  leadership-class computers and by commodity clusters, and
  theoretical advances enabling extrapolations to be performed from
  the quark masses employed in the computation to the physical up- and
  down-quark masses.  In this talk, I will show the impact of these
  advances on hadron structure, and provide a roadmap for future
  calculations.  The layout of the rest of this talk is as follows.
  In the next section, I will describe how hadron-structure
  calculations proceed in lattice QCD.  I will then review benchmark
  calculations, such as that of the nucleon's electromagnetic form factors.
  Section~\ref{sec:gpd} will describe calculations of moments of
  Generalized Parton Discributions (GPDs) and the impact such
  calculations are having on our understanding of hadron structure.  I
  will conclude with prospects for future calculations.

\section{Anatomy of a nucleon-structure calculation}
\label{sec:basics}
Hadron structure is expressed through quantities as the
electromagnetic form factors, describing the distribution of charge
and currents within a hadron, and the polarized and unpolarized
parton structure functions, describing how the longitudinal momentum
fraction and spin is apportioned amongst the constituents.  Recently,
Generalized Parton Distributions have been
introduced\cite{Mueller:1998fv,Ji:1996nm,Radyushkin:1997ki},
encompassing both of the concepts above, but allowing a
three-dimensional picture of the nucleon to be constructed.

GPDs are expressed as the matrix elements of light-cone correlation
functions ${\cal O}_{\Gamma}(x)$:
\begin{equation}
{\cal O}_{\Gamma}(x) = \int \frac{ d \lambda}{4\pi} e^{i \lambda \, x} \bar{q}
(\frac{ - \lambda n}{2}) \Gamma {\cal P} e^{- i g
  \int_{-\lambda/2}^{\lambda/2} d \alpha n \cdot A(\alpha n)}
q(\frac{\lambda}{2})\label{eq:gpd}
\end{equation}
where $n$ is a light-cone vector, the parallel-transporter is
necessary to ensure gauge invariance, and the flavour indices on the
quarks are suppressed.  The familiar quark polarized and unpolarized
parton structure functions are then forward matrix elements of this
operator, with $\Gamma = \gamma_{\mu}$ or $\gamma_{\mu} \gamma_5$ for
the unpolarized and polarized distributions, respectively; the GPDs
correspond to matrix elements in the off-forward direction, with
different momenta for the incoming and outgoing partons.

The use of a Euclidean lattice precludes the measurement of these
matrix elements directly.  Instead with appeal to the Operator Product
Expansion to expand the operator ${\cal O}(x)$ about the light cone,
yielding a set of local operators which can be measured on a Euclidean
lattice, and furthermore analytically continued to Minkowski space.
In particular, the moments, with respect to Bjorken-$x$, of the
quark structure functions are obtained in terms of the
forward matrix elements
\begin{equation}
{\cal O}_{\Gamma}^{\{ \mu_1\dots \mu_n \}} =   \bar{q} (\gamma_5) \gamma^{\{
  \mu_1}  i D^{\mu_2} \dots
D^{\mu_n \}} q.\label{eq:operator}
\end{equation}
Specifically, for the unpolarized distribution for a nucleon carrying
momentum $\vec{p}$, we have
\begin{equation}
\langle \vec{p} | {\cal O}_q^{\mu_1 \dots \mu_n} \mid \vec{p}
\rangle \longrightarrow
\int _0^1 dx \, x^{n-1} q(x),
\end{equation}
where I have suppressed the Lorentz structure.  

\begin{figure*}
\centering
\includegraphics[width=400pt]{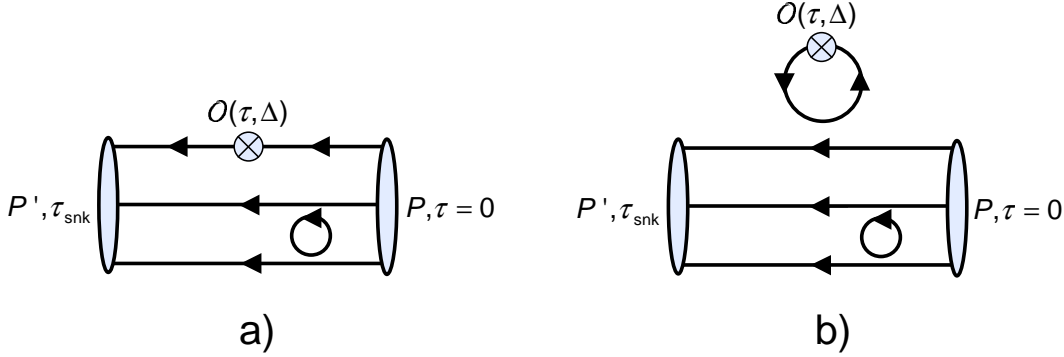}
\caption{The left- and right-hand panels show the connected and
  disconnected contributions to the three-point function of
  \eqn{eq:threept}.\label{fig:threept}}
\end{figure*}

We have seen in the talk of Christof Gattringer how the spectrum is
obtained in lattice QCD through the measurement of so-called two-point
functions representing the Euclidean correlators of two operators that
interpolate between a state and the vacuum.  Nucleon structure is
investigated through the measurement of \textit{three-point}
functions, illustrated in Figure~\ref{fig:threept}, representing the
Euclidean correlation functions of three operators:
\begin{eqnarray}
\lefteqn{C_{\rm 3pt}(\tsnk, \tau; \vec{p}, \vec{q}) = }\nonumber \\
& & \sum_{\vec{x}, \vec{y}}
e^{-i \vec{p} \cdot \vec{x} -i \vec{q}\cdot \vec{y}}
\langle 0 | J(\vec{x}, \tsnk) {\cal O}(\vec{y},\tau) J^{\dagger} (0) | 0 \rangle,
\label{eq:threept}
\end{eqnarray}
where $J$ is an operator that interpolates between the state and the
vacuum. Note that the disconnected contribution in the right-hand panel of
Figure~\ref{fig:threept} is far more computationally demanding that
the connected piece; for most of the following, I will emphasise
the calculation of flavour non-singlet, or isovector, quantities, for
which the disconnected piece vanishes.  

Following~\cite{gattringer}, we insert complete sets of
states between the operators in \eqn{eq:threept}; the time-slice sum
projects onto states of definite momentum, and we obtain:
\begin{eqnarray}
\lefteqn{C_{\rm 3pt}(\tsnk, \tau; \vec{p}, \vec{q}) = \sum_{n=1,\dots} e^{-E_n(\vec{p}) (\tsnk -
    \tau)} e^{- E_n(\vec{p} + \vec{q}) \tau } \times } \nonumber \\
& & \langle 0 | J(0) | n,
\vec{p} \rangle
\langle n, \vec{p} | {\cal O}(0) | n, \vec{p} +
\vec{q} \rangle \langle n, \vec{p} + \vec{q} | J^{\dagger}(0) |
0 \rangle.
\end{eqnarray}
For sufficiently large $\tau, \tsnk - \tau$, $C_{\rm 3pt}$ is dominated by the
lowest-lying state, and we can obtain $\langle n = 1, \vec{p} | {\cal O} |
n= 1, \vec{p} \rangle$ after elimination of the vacuum-to-state matrix
elements from the corresponding two-point functions.

Computations of hadron structure are being performed by several
groups, using a variety of fermion and gauge-field discretisations.
The common theme is calculations at increasingly light values of the
pion mass, enabling lattice calculations, together with chiral
effective theory, to provide insight into physics at the physical
quark masses; recent reviews are contained in
ref.~\cite{Hagler:2007hu,Orginos:2006zz}.  During the remainder of
this talk, I will in large part quote results from the LHP
Collaboration.  We employ a hybrid approach, using lattices generated
using $2 + 1$ flavours of Asqtad quarks, corresponding to degenerate
$u/d$ and strange, by the MILC Collaboration\cite{Bernard:2001av}, but
with domain-wall fermions (DWF) for the valence quarks, computed on
HYP-smeared lattices\cite{Hasenfratz:2001hp}.  The generation of
lattices using Asqtad fermions is very computationally efficient,
whilst the DWF valence quarks have the desired chiral-symmetry
properties, avoiding the much operator mixing and simplifying the
matching of our results to the continuum.  Whilst this approach
violates unitarity at finite lattice spacing, we expect this approach
to yield the correct theory in the continuum
limit\cite{Sharpe:2006re}.  In this talk, I will present results with
pion masses down to around 360~MeV.

\section{Electromagnetic Form Factors}
\begin{figure}
\centering
\includegraphics[width=200pt]{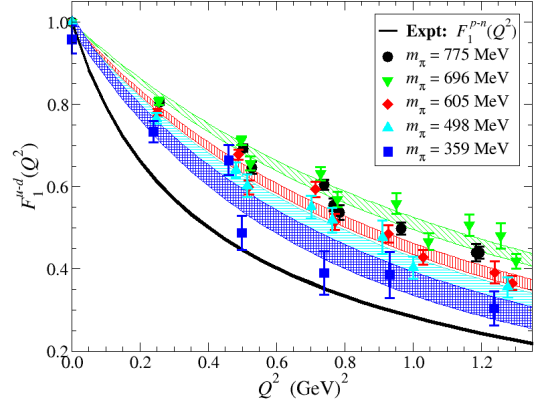}
\caption{The points show measurements of the isovector $F_1$ form
  factor at five values of the pion mass; the coloured bands are uncertainties
  from a pole-dominance fit to the data.  The solid line is a
  parametrisation of the experimental
  data\protect\cite{Kelly:2004hm}.\label{fig:F1}}
\end{figure}

These describe the distribution of charge and current within the
nucleon, and remain the subject of intense experimental and
theoretical interest.  They are the most straightforward quantity to
measure in a lattice calculation, corresponding to the matrix element
of the electromagnetic current $V_{\mu} = \bar{q} \gamma_\mu q$.
Figure~\ref{fig:F1} shows the form factor $F_1(Q^2)$ determined by the
LHP Collaboration\cite{Edwards:2006qx}, together with a
parametrisation of the isovector experimental
data\cite{Kelly:2004hm}.  The data show the approach to the
experimental parametrisation at decreasing values of the pion mass
used in the lattice calculation.

\begin{figure*}
\vspace{0.5cm}
\centering
\includegraphics[width=200pt]{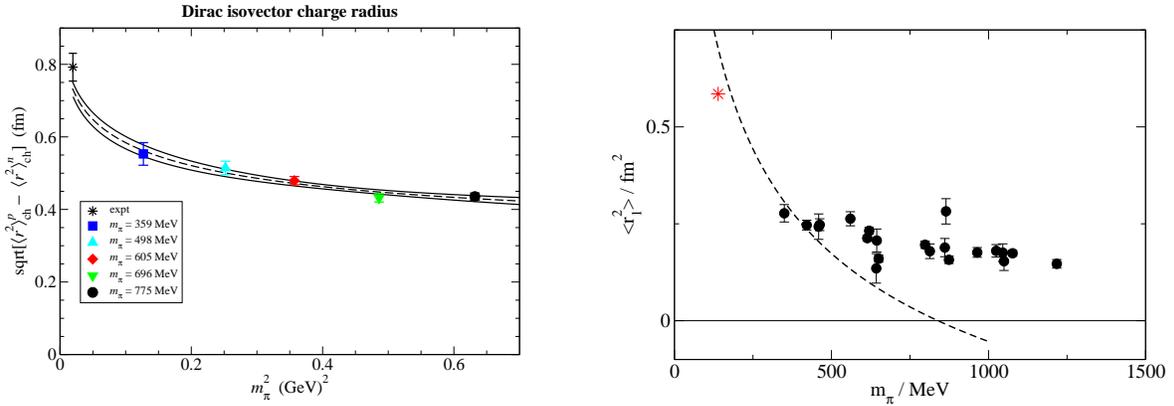}\hspace{1cm}
\includegraphics[width=210pt]{rv1Q.eps}
\caption{The left-hand panel shows the isovector charge radius using a
  mixed Asqtad/DWF calculation\cite{Edwards:2006qx}, together with the
  chiral behavious using a finite-range
  regulator\cite{Leinweber:2001ui}.  The right-hand panel shows the
  mean-square charge radius using two flavours of Wilson
  fermions\cite{Gockeler:2007hj}, together with the chiral behaviour
  using the small-scale
  expansion\cite{Gockeler:2003ay}.\label{fig:charge}}
\end{figure*}

The slope of the form factor at $Q^2 = 0$ is related to the isovector
charge radius $\langle r^2 \rangle^{1/2}$.  The left-hand panel in
Figure~\ref{fig:charge} shows the charge radius obtained from the
calculation above.  A naive linear fit in $m_\pi^2$ would clear be far
below the experimental points, illustrating the need to correctly
describe the non-analytic behaviour in the approach to the physical
light-quark masses.  The curve shows the chiral extrapolation of the
charge radius, using the finite-range
regulator\cite{Leinweber:2001ui}.  The right-hand panel shows the
isovector mean square charge radius obtained by the QCDSF/UKQCD
Collaborations, using two flavours of Wilson fermions, with pion
masses down to around $350~{\rm MeV}$; the line is the chiral
behaviour in the small-scale expansion.\cite{Gockeler:2003ay}

\section{Generalized Parton Distributions}
\label{sec:gpd}
The electromagnetic form factors described above are particular cases
of Generalized Parton Distributions.  There are two GPDs corresponding
to the choice $\Gamma = \gamma \cdot n$ in \eqn{eq:gpd}, $H(x,\xi,t)$
and $E(x,\xi,t)$, and a further two corresponding to the $\Gamma =
\gamma \cdot n \gamma_5$ in \eqn{eq:gpd}, $\tilde{H}(x,\xi,t)$
and $\tilde{E}(x,\xi,t)$.  The invariants are:
\begin{eqnarray}
t & = & - \Delta^2 = -(P - P')^2 \nonumber \\
\xi & = & -n \cdot \Delta /2
\end{eqnarray}
where $n_\mu$ is a light-cone vector, and $P',P$ are the four-momenta
of the incoming and outgoing states.  Once again, we appeal to the
operator product expansion to obtain the matrix elements of local
operators, and both the form factors described above and the familiar
parton distributions emerge as special cases:
\begin{eqnarray}
H(x,0,0) & = & q(x)\\
\tilde{H}(x,0,0) & = & \Delta q(x),
\end{eqnarray}
and
\begin{equation}
\int_{-1}^{1} dx \, H(x,\xi,t) = F_1(t).
\end{equation}
Note here that the $x$ runs from -1 to 1, corresponding to antiquark
and quark momentum fraction.

\begin{figure}
\centering
\includegraphics[width=200pt,angle=-90]{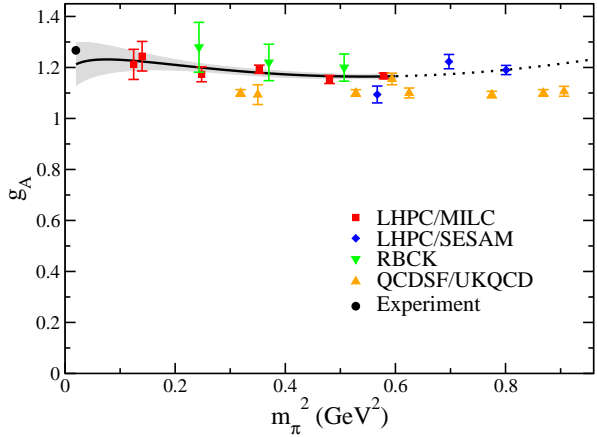}
\caption{The red squares show LHPC/MILC data for the axial-vector
  charge $g_A$; the two data points at the lowest pion mass correspond
  to to different volumes.  The remaining lattice points are taken
  from QCDSF/UKQCD\cite{Khan:2004vw}, RBCK\cite{Ohta:2004mg} and
  SESAM\cite{Dolgov:2002zm}.\label{fig:ga}}
\end{figure}

Before proceeding to discuss the new insights enabled by the study of
GPDs, we begin by discussing the polarized and unpolarized
distributions as a means of benchmarking our calculation.  A
fundamental measure of QCD is the nucleon's axial-vector charge $g_A$,
corresponding to the choice $\Gamma = \gamma_5$; it has additional
importance in its r\^{o}le as a fundamental low-energy constant of the
theory that appears in the chiral expansion of other quantities.
Figure~\ref{fig:ga} shows a calculation of the axial-vector charge by
the LHP collaboration, together with the chiral extrapolation to the
physical quark masses\cite{Edwards:2005ym}.  The consistency between the extrapolated
lattice data and the experimental value is very encouraging, though
the chiral behaviour for this quantity is very mild.

\begin{figure*}
\centering
\includegraphics[width=200pt]{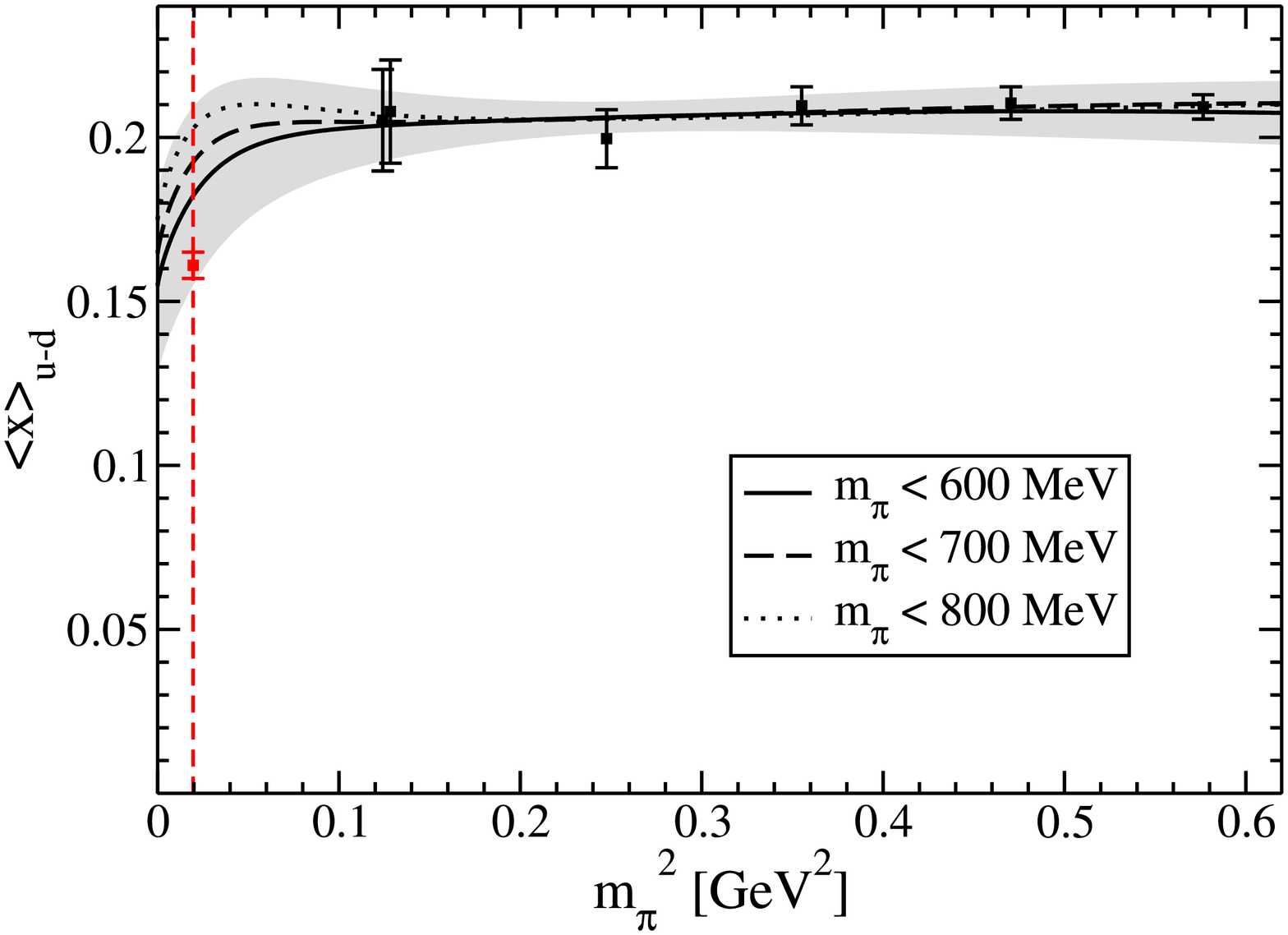}\hspace{1cm}
\includegraphics[width=200pt]{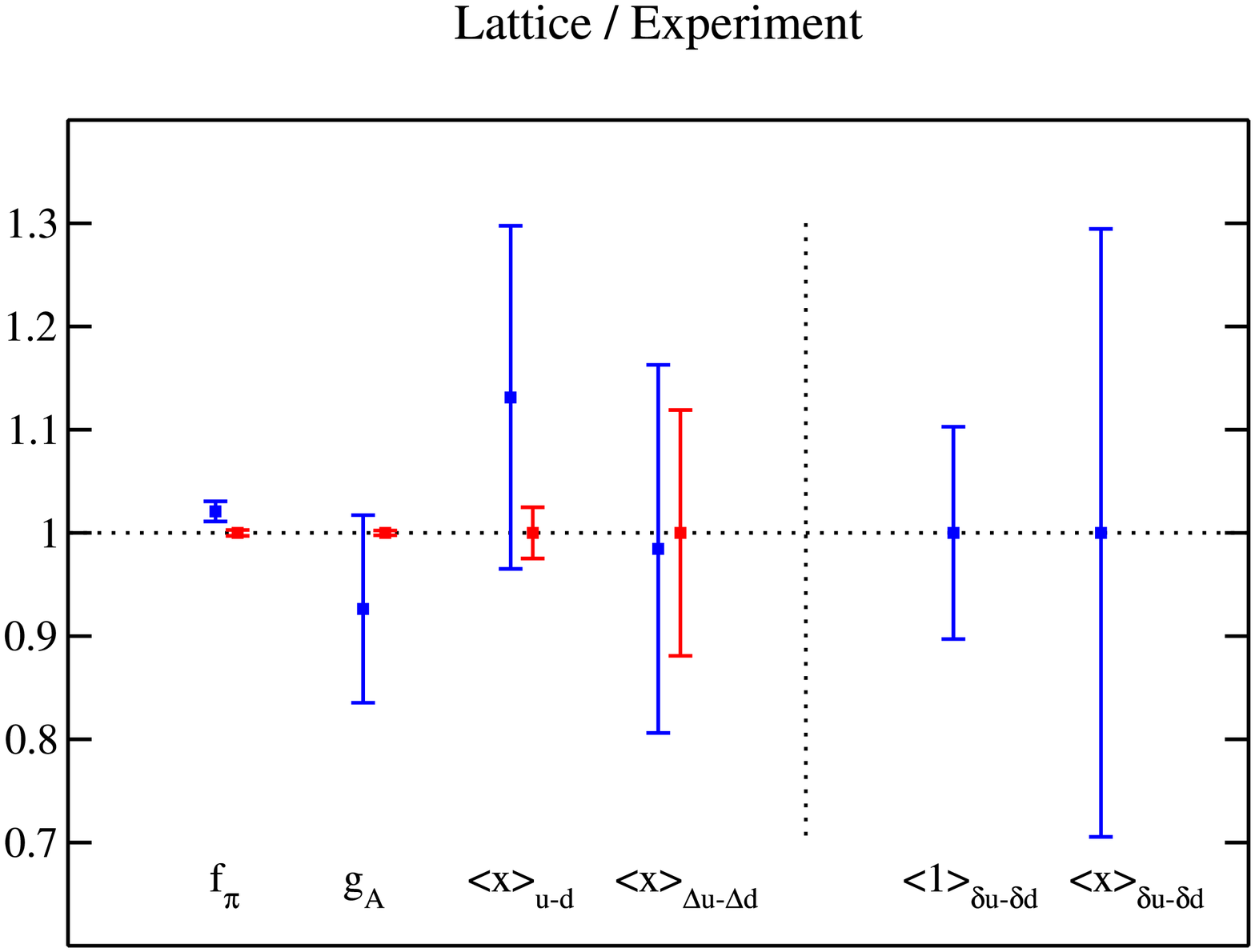}
\caption{The left-hand figure shows the momentum fraction carried by
  the valence quarks in a nucleon, together with the chiral
  extrapolation to the physical quark masses.  The right-hand figure
  is a summary shows the lattice values at the physical pion masses
  for a range of benchmark quantities, normalized to the experimental
  value where known (blue)\cite{Renner:2007pb}; the errors on the corresponding
  experimental measurements are shown in red.\label{fig:benchmark}}
\end{figure*}

The non-analytic behaviour with decreasing quark mass is more strongly
exhibited in the unpolarized distributions, and in particular in the
calculation of the momentum fraction carried by the valence quarks in
the nucleon.  An analysis of the forward matrix elements for both the
unpolarized and polarized distributions, as well as for the
transversity, has been performed using a self-consistent chiral
expansion\cite{Renner:2007pb}; data for the non-singlet momentum
fraction, and a summary of the results for a range of benchmark
quantities, is shown in Figure~\ref{fig:benchmark}.

\subsection{New insights: origin of nucleon spin}
The excitement and signficance of GPDs arises from the new insights
they can provide into hadron structure, spurring new experimental
initiatives such as the 12 GeV upgrade at Jefferson Laboratory.  There
is thus a rigorous effort both by the LHP Collaboration in a hybrid
DWF/Asqtad approach\cite{Hagler:2007xi,Renner:2007pb}, and by the
QCDSF/UKQCD Collaboration using two flavours of dynamical, improved
Wilson quarks\cite{Ohtani:2007sb}, to measure moments of GPDs.

\begin{figure*}
\centering
\parbox{3in}{\includegraphics[width=150pt,angle=-90,clip=true]{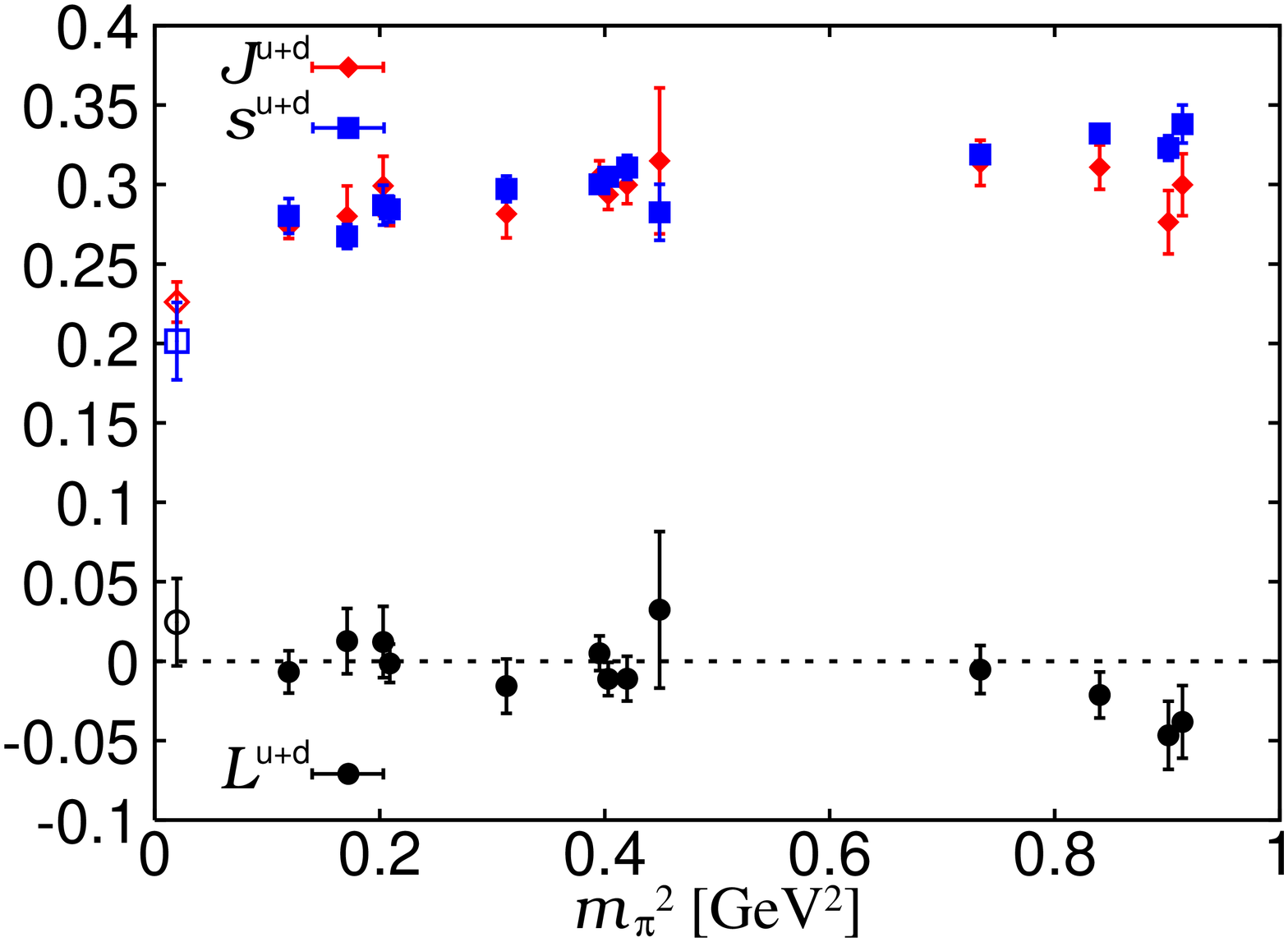}}\hspace{1cm}
\parbox{3in}{\includegraphics[width=210pt, clip=true]{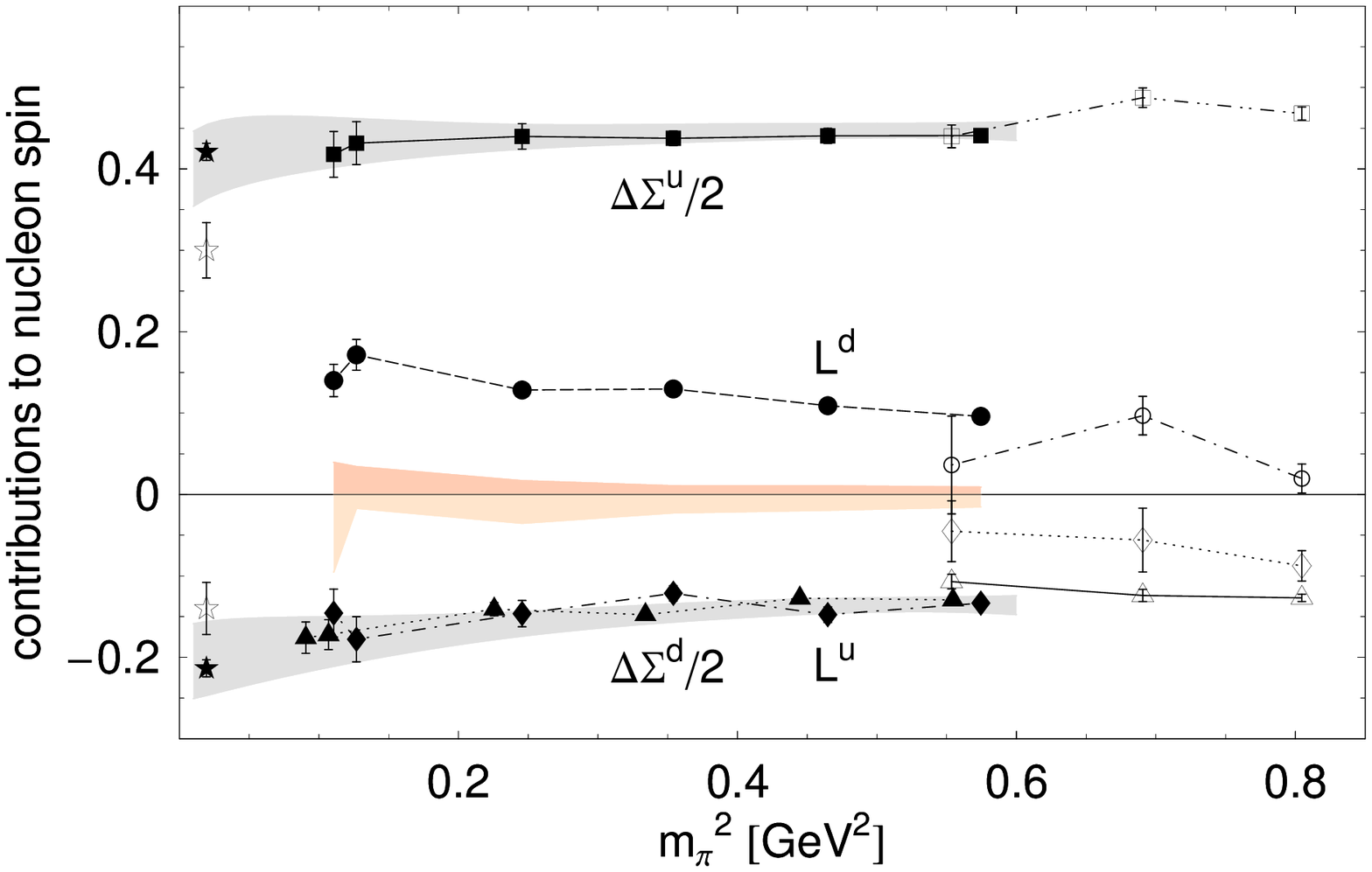}}
\caption{The left-hand plot shows the spin carried by the quarks in
  the nucleon, the total angular momentum carried by the quarks, and
  the orbital angular momentum, in a calculation with two flavours of
  improved Wilson quarks\cite{Ohtani:2007sb}.  The right-hand plot
  shows the decomposition of the spin between the quark flavours in a
  hybrid Asqtad/DWF approach\cite{Hagler:2007xi}, together with an
  experimental determination of the spin from
  HERMES\cite{Airapetian:2007mh,Ackerstaff:1999ey}\label{fig:orbital}
  }
\end{figure*}

An salient example of the new insights facilitated through the study
of GPDs is how the spin of the nucleon is distributed amongst its
constituents, and in particular the r\^{o}le of orbital angular
momentum of the quarks.  The spin carried by the quarks in a nucleon
has long been the pursuit of lattice calculations, but no direct
gauge-invariant definition of the orbital angular momentum appeared
possible, though a new proposal has been presented
here\cite{Chen:2007wb}.  However, it was realized that the total
angular momentum carried by the quarks within a nucleon could be
related to moments of GPDs through\cite{Ji:1996ek}
\begin{equation}
J_q = \frac{1}{2} \int dx \, x \left( H(x,\xi,0) + E(x, \xi,0) \right).
\end{equation}
A pioneering effort to measure the quark orbital angular momentum was
performed in the quenched approximation to QCD\cite{Mathur:1999uf};
Figure~\ref{fig:orbital} shows more recent calculations by the LHPC and
QCDSF/UKQCD Collaborations.  Note that the calculations include only
the connected pieces.  None-the-less, they suggest that the total
orbital angular momentum carried by the quarks within a nucleon is
small, but that the orbital angular momentum carried by the individual
quark flavours may be substantial.

\begin{figure}
\centering
\parbox{200pt}{\vspace{1cm}\includegraphics[width=200pt,clip=yes]{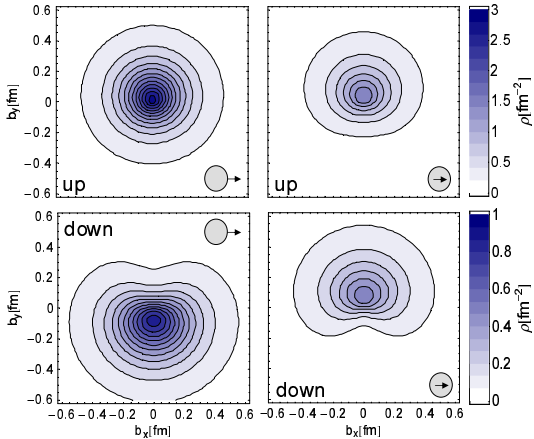}}
\caption{Lowest moment of the densities of unpolarized quarks in a
  transversely polarized nucleon (left), and transversely polarized
  quarks in an unpolarized nucleon (right)\cite{Gockeler:2006zu}.  The
  upper and lower figures correspond to the up and down quarks
  respectively.\label{fig:densities}}
\end{figure}

Intimately related to the question of the origin of orbital angular
momentum in the nucleon are the so-called transverse spin densities,
corresponding to the choice $\Gamma = \sigma^{\mu\nu}\gamma_5$ in
\eqn{eq:gpd}.  These can be studied in lattice QCD, complementing the
experimental effort at investigating Tranverse-momentum-dependent
PDFs.  Figure~\ref{fig:densities} shows the lowest moment of the
densities for an unpolarised quark in an transversely polarized
nucleon, and for transversely polarized quarks in an unpolarized
nucleon\cite{Gockeler:2006zu}.  The right-hand plots reveal the strong
distortion in the spin densities for a transversely polarized quark in
an unpolarized nucleon, and are suggestive of a negative Boer-Mulders
function\cite{Boer:1997nt}.

\section{Structure of resonances}
\begin{figure*}
\centering
\includegraphics[width=200pt]{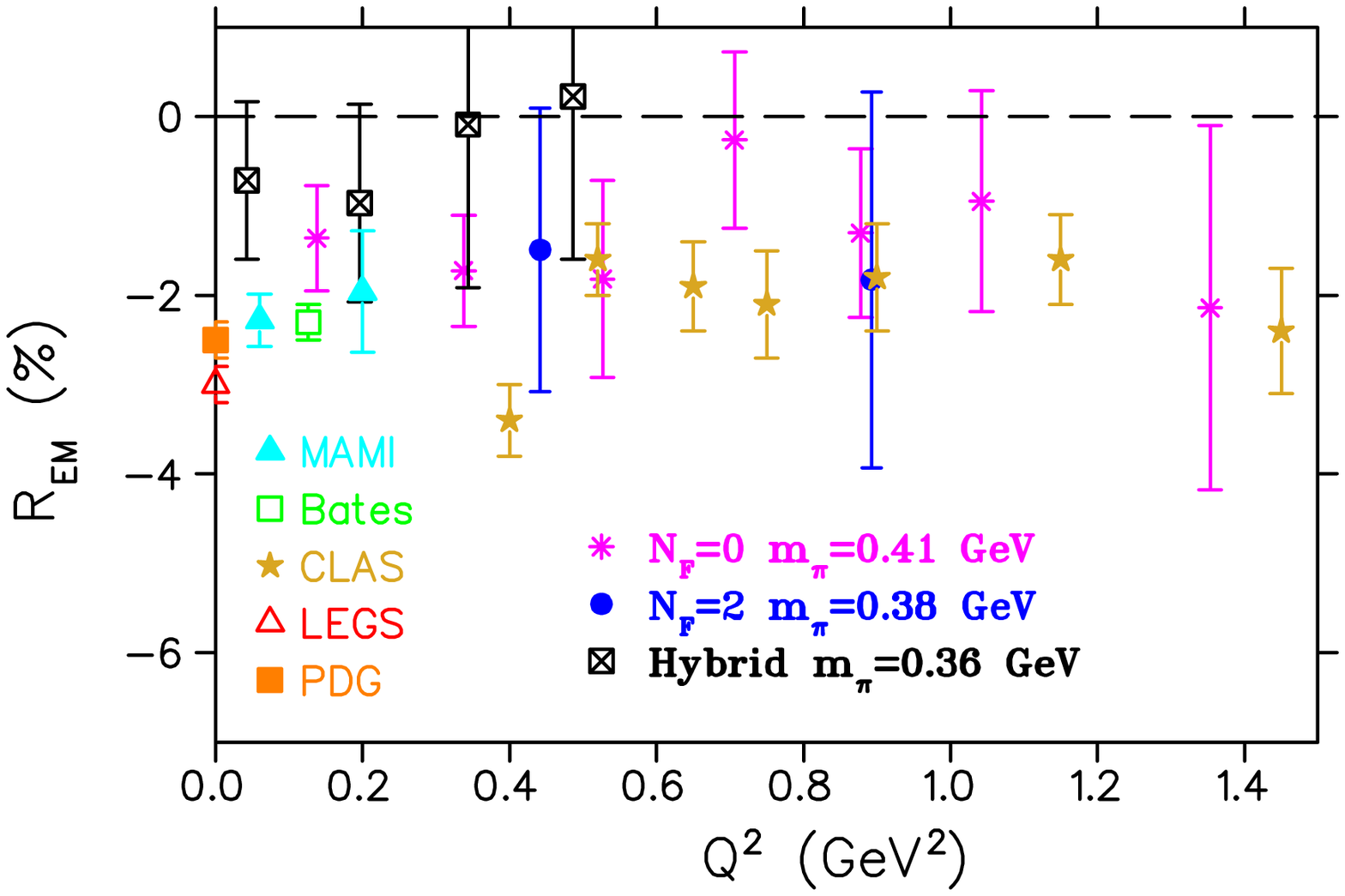}
\includegraphics[width=200pt]{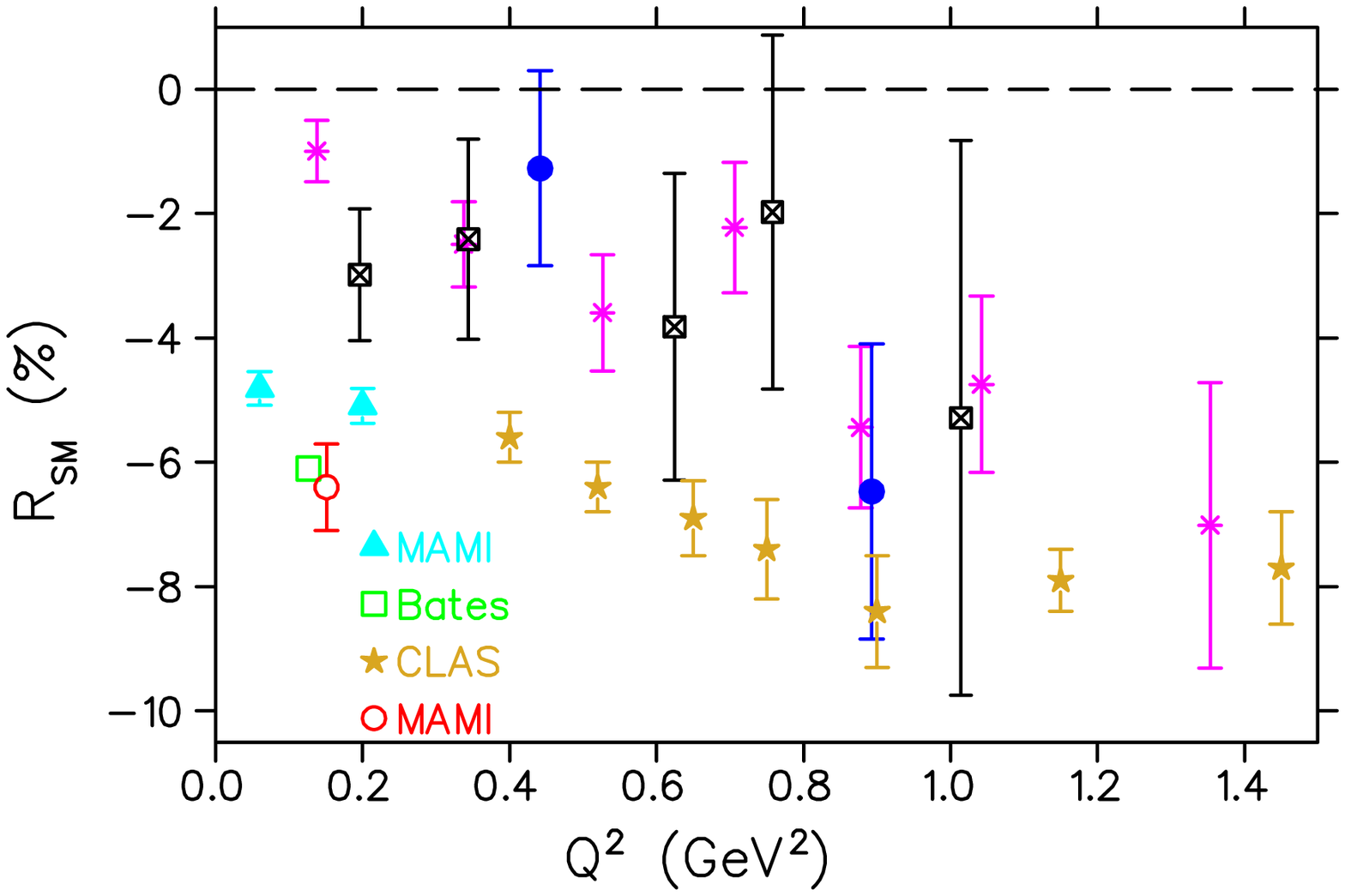}
\caption{The left- and right-hand plots show the ratios ${\rm EMR} =
  -{\cal G}_{E2}(Q^2)/{\cal G}_{M1}(Q^2)$ and ${\rm CMR} =
  -(|\vec{q}|/2 m_{\Delta}) {\cal G}_{C2}(q^2)/{\cal
    G}_{M1}(q^2)$\cite{Alexandrou:2007dt}, using both quenched and
  unquenched Wilson fermion actions, and using the Hybrid Asqtad/DWF
  approach.  Also shown are experimental measurements.\label{fig:delta}}
\end{figure*}

I have emphasised in this talk the structure of the nucleon, and have
so far said little of the structure of nucleon resonances.  The one
resonance that has been extensively studied in lattice QCD is the
$\Delta$, and indeed the mass of the $\Delta$ is a benchmark
calculation in the hadron spectrum.  The $\Delta$ decays to $N\pi$ in
a $P$-wave, and the finite spatial extent used in our lattices
provides a momentum threshold forbidding such decays in most
calculations at current volumes and pion masses.  Nevertheless,
extensive insight into the structure of QCD has been revealed, in
particular through lattice calculations of $\Delta$ to $N$
electromagnetic transition form factors.  In contrast to the
electromagnetic form factors discussed earlier, these are sensitive to
a quadrupole moment in the nucleon and hence deviations from spherical
symmetry.  Recently, there has been a calculation of the ratio of
electric-to-magnetic (EMR) and Coulomb-to-magnetic (CMR) transition
form factor ratios, shown in Figure~\ref{fig:delta}, with sufficient
accrucacy to exclude a zero value\cite{Alexandrou:2007dt}; a non-zero
value is indicative of deformations in the $N-\Delta$ system.

Given a sufficient basis of interpolating operators, the determination
of the transitions to radial excitations will be feasible; we have
heard in this workshop the success at extracting the electromagnetic
properties of the Roper resonance from the CLAS data at Jefferson
Laboratory\cite{Aznauryan:2007pi}. Indeed, the first attempts at studying the
Roper in lattice QCD are in progress\cite{hwlin:aps}.

\section{Conclusions and future prospects}
Theoretical and computational advances are enabling lattice QCD to
have a vital impact on our understanding of hadron structure.  Future
challenges are two-fold: to proceed to values of the light-quark
masses closer to the physical quark masses so to provide stronger
constraints on the chiral extrapolations, and to delineate the
contributions of the individual quark flavours.  We heard in the talk
of Christof Gattringer the successes at resolving the excited-state
nucleon masses; the first attempts to study their properties should
appear shortly.

{\bfseries Acknowledgements:} This work was supported by DOE contract
DE-AC05-06OR23177 under which the Jefferson Science Associates, LLC
operates the Thomas Jefferson National Accelerator
Facility.  I am grateful for collaboration with my colleagues in LHPC:
D.B.~Renner, J.~Bratt, R.G.~Edwards, M.~Engelhardt, G.~Fleming,
Ph.\ H\"{a}gler, B.~Musch, J.W.~Negele, K.~Orginos, A.V.~Pochinsky, and
W.~Schroers

 \bibliographystyle{epj}
 \bibliography{richards}
\end{document}